\documentclass[a4paper,12pt]{article}

\usepackage{amsmath,amssymb}
\usepackage{graphicx}
\usepackage{hyperref}

\usepackage{url}
\usepackage{epsfig}
\usepackage{amsmath}
\usepackage{amssymb}
\usepackage{amsfonts}
\usepackage{bbm}

\usepackage[footnotesize]{caption}
\usepackage{graphicx}
\usepackage{mathrsfs}
\usepackage[font=scriptsize]{subcaption}
\usepackage{color}
\usepackage{comment}
\usepackage{mathtools}
\usepackage{braket}
\usepackage{cite}
\usepackage{hyperref}
\usepackage{multirow}
\usepackage{relsize}
\usepackage{fullpage}
\usepackage{makecell}

\usepackage{bbold}

\usepackage{color}

\newcommand{\ii}{\mathrm{i}}

\newcommand{\CP}{$\mathcal{CP}$\,}

\newcommand{\Tr}{\mathrm{Tr}\,}

\newcommand{\Id}{\mathbb{1}\,}

\newcommand{\secheadmath}[1]{\texorpdfstring{$#1$}{TEXT}}

\begin{document}

\begin{titlepage}
\begin{center}
\textbf{\Large Invariant approach to \CP in unbroken $\Delta(27)$} \\[12mm]
Gustavo~C.~Branco$^{\dagger}$
\footnote{E-mail: \texttt{gbranco@tecnico.ulisboa.pt}},
Ivo~de~Medeiros~Varzielas$^{\star}$
\footnote{E-mail: \texttt{ivo.de@soton.ac.uk}},
Stephen~F.~King$^{\star}$
\footnote{E-mail: \texttt{king@soton.ac.uk}}
\\[-2mm]

\end{center}
\vspace*{0.50cm}
\centerline{$^{\dagger}$ \it
Centro de F{\'\i}sica Te\'orica de Part{\'\i}culas - CFTP,}
\centerline{\it Instituto Superior T\'ecnico - IST, Universidade de Lisboa - UL,}
\centerline{\it Avenida Rovisco Pais, 1049-001 Lisboa, Portugal}
\vspace*{0.2cm}
\centerline{$^{\star}$ \it
School of Physics and Astronomy, University of Southampton,}
\centerline{\it
SO17 1BJ Southampton, United Kingdom }
\vspace*{1.20cm}

\begin{abstract}
{\footnotesize
\noindent The invariant approach is a powerful method for studying \CP violation for specific Lagrangians. 
The method is particularly useful for dealing with discrete family symmetries.
We focus on the \CP properties of unbroken $\Delta(27)$ invariant Lagrangians with Yukawa-like terms,
which proves to be a rich framework, with distinct aspects of \CP, making it an ideal group to investigate with the invariant approach. 
We classify Lagrangians  depending on the number of fields transforming as irreducible triplet representations of $\Delta(27)$. 
For each case, we construct \CP-odd weak basis invariants and use them to discuss
the respective \CP properties.
We find that \CP violation is sensitive to the number and type of $\Delta(27)$ representations.
}
\end{abstract}

\end{titlepage}

\section{Introduction}

The origin and nature of \CP and its violation remains a mystery both within and beyond the Standard Model (SM).
In addressing the question of \CP it was observed some time ago that phases which appear in the Yukawa matrices
for example are not robust indicators of \CP violation since their appearance is dependent on the choice of basis.
On the other hand, physical \CP violating observables only depend on particular combinations
of Yukawa matrices which are invariant under different choices of basis. Such weak basis invariants,
which have the property that they are zero if \CP is conserved and non-zero if \CP is violated 
therefore provide unambiguous signals of \CP violation which are closely related to experimentally measurable quantities. The use of such \CP-odd weak-basis invariants (CPIs), rather than particular phases in a given basis, is generally referred to as the Invariant Approach (IA) to \CP violation.

In the IA to \CP violation \cite{Bernabeu:1986fc}, one starts by separating the full Lagrangian of the theory in two parts, one denoted $\mathcal{L_{CP}}$  that is known to conserve \CP, typically the kinetic terms and pure gauge interactions \cite{Grimus:1995zi} \footnote{
The use of \CP-like transformations that include a family symmetry transformation was introduced in \cite{Ecker:1981wv}, in an attempt to obtain a connection between quark masses and mixing angles.}, and the remaining Lagrangian, denoted $\mathcal{L}_{rem.}$. The crucial point is that $\mathcal{L_{CP}}$ allows for many different \CP transformations and as a result, \CP is violated if and only if none of these \CP transformations leaves $\mathcal{L}_{rem.}$ invariant.
In the case of the SM, $\mathcal{L_{CP}}$  includes the gauge interactions and the kinetic energy terms, while the relevant components of $\mathcal{L}_{rem.}$ are the Yukawa interactions. Using the IA, one can readily derive \cite{Bernabeu:1986fc} some specific conditions that the Yukawa couplings have to satisfy in order to have \CP invariance. It is well known that the Yukawa couplings in the SM have a large redundancy which results from the freedom that one has to make redefinitions of the fermion fields which leave the gauge interactions invariant but change e.g. the quark Yukawa couplings $Y_u$, $Y_d$ without changing the Physics. The great advantage of the IA is that it allows one to derive CPIs which, if non-vanishing, imply \CP violation. In the SM, it has been shown \cite{Bernabeu:1986fc} that the relevant CPI is
$\Tr [H_u , H_d]^ 3$,
where we define the Hermitian combinations $H_u \equiv Y_u Y_u^\dagger$ and $H_d \equiv Y_d Y_d^\dagger$. For the 3 fermion generation case this CPI leads to the Jarlskog invariant \cite{Jarlskog:1985ht}.
The IA can be applied to any extension of the SM, in particular to extensions of the SM with Majorana neutrinos \cite{Branco:1986gr}.

It should be emphasised that the IA not only enables one to verify whether a given Lagrangian violates \CP, but also provides an idea of how suppressed \CP violation might be. A notable example is the possibility of showing why \CP in the SM is too small to generate the baryon asymmetry of the Universe (BAU). One simply observes that the dimensionless number
$\Tr [M_u M_u^\dagger , M_d M_d^\dagger]^ 3 / v^{12}$ is of order $10^{-20}$, where we used the Hermitian quark mass matrices and $v = 246$ GeV denotes the scale of electroweak symmetry breaking. This dimensionless number should be compared to the size of observed BAU, $n_B/n_\gamma \simeq 10^{-10}$ \cite{Ade:2015xua}.
The IA, leading to basis invariant quantities, also identifies what combination of parameters are physical such that, e.g. there is no need to count how many phases can be eliminated through rephasing, which can be laborious in complicated Lagrangian, and specially in the presence of family symmetries.

Recently  \cite{Branco:2015hea} the use of CPIs, valid for any choice of \CP transformation, was advocated as a powerful approach to studying specific models of \CP violation in the presence of discrete family symmetries. 
Examples based on $A_4$ and $\Delta(27)$ family symmetries were discussed and it was shown 
how to obtain several known results in the literature. In addition, 
the IA was used to identify how explicit (rather than spontaneous) \CP violation arises, which is geometrical in nature, i.e. persisting for arbitrary couplings in the Lagrangian.

Here we intend both to further highlight the usefulness of the IA in dealing with discrete family symmetries and also to systematically explore the \CP properties of $\Delta(27)$. By using the IA, we are able to construct CPIs independently of the specific group and need to consider the group details only to compute coupling matrices by using the respective Clebsch-Gordan coefficients in any particular basis. By combining the coupling matrices with the CPIs, basis-independent quantities are obtained which indicate if there is \CP violation.

In this paper we explore in depth the \CP properties of unbroken $\Delta(27)$ invariant Lagrangians using the IA as outlined in \cite{Branco:2015hea} (see also the proceedings \cite{Varzielas:2015fxa}). The method is based on \cite{Bernabeu:1986fc}.
We focus on $\Delta(27)$ since it involves many features which may be encountered in 
more general discrete groups, such as complex representations and multiple non-trivial 
singlet representations. It therefore constitutes a rich playground for exploring the IA in the case of discrete family symmetries.
Although the cases discussed do not represent realistic models, since the $\Delta(27)$ is unbroken,
the work here lays the foundation for future models based on spontaneously broken $\Delta(27)$.

Following the IA we consider several cases that highlight how the \CP properties depend both on the field content and on the type of contractions considered (which may be controlled e.g. by additional symmetries, even though here we don't always consider those explicitly).
We focus on tri-linears terms, which we refer to as Yukawa-like couplings, keeping in mind that most (but not all) cases considered are meant as fermion-fermion-scalar terms.
We start by considering Lagrangians with just $\Delta(27)$ singlets where the IA identifies the relative phases that are physical, and then concentrate on Yukawa-like terms involving a triplet, an anti-triplet, and a singlet.

When the Yukawa-like couplings are between triplet, anti-triplet and singlet, we study cases with a single independent triplet, with independent triplet and anti-triplet, and with three or more independent 3-dimensional irreducible representations. Some of the cases considered are similar to adding $\Delta(27)$ to a type II 2 Higgs doublet model (2HDM) \footnote{See e.g. \cite{Branco:2011iw} for a review.} or $N$ Higgs doublet model (NHDM).
For each of these frameworks the IA allows to identify if the Lagrangian has the possibility to explicitly violate \CP, how this depends on how many different $\Delta(27)$ singlets are coupled, and what CPIs are relevant and non-vanishing when there is \CP violation.
This serves to further illustrate the convenience and power of the IA for the study of \CP properties of specific Lagrangians.

The layout of the remainder of this paper is as follows. In section~\ref{IA} we briefly review the IA to 
\CP in family symmetry models. We continue with the group theory of $\Delta(27)$ in section~\ref{D27}. Section~\ref{singlets} considers just singlets. In section~\ref{one} the field content includes one triplet. Two triplets are considered in detail in section~\ref{2triplets}, where we differentiate also based on the number and type of singlets present. We generalise to three triplets in section~\ref{three} and to four and more triplets in appendix \ref{sec:4more}. For comparison with the IA, we present some examples of specific \CP matrices in appendix \ref{spe}.  Finally we conclude in section~\ref{Conclusions}.

\section{Invariant approach to \CP in family symmetry models}
\label{IA}

As mentioned in the introduction, the IA as outlined in \cite{Branco:2015hea,Varzielas:2015fxa} is based on \cite{Bernabeu:1986fc}, where to study the \CP properties of a given Lagrangian one starts by splitting it
\begin{equation}
\mathcal{L}=\mathcal{L_{CP}}+\mathcal{L}_{rem.} \,,
\label{L}
\end{equation}
where $\mathcal{L_{CP}}$ denotes the part that is known to conserve \CP (kinetic terms and gauge interactions, as pure gauge interactions conserve \CP  \cite{Grimus:1995zi}). $\mathcal{L}_{rem.}$ includes non-gauge interactions such as the Yukawa couplings.
A review of how the IA is applied to the Standard Model (SM) lepton sector can be found in \cite{Branco:2015hea}, which also includes its application to a model of spontaneously broken $A_4$ and to a model of $\Delta(27)$ which features explicit geometrical \CP violation.

In this paper we study many different Lagrangians invariant under unbroken $\Delta(27)$, relying on the IA. As pointed out in \cite{Branco:2015hea}, the presence of a family symmetry does not change the most general \CP transformation which leaves invariant $\mathcal{L_{CP}}$ - these are the kinetic terms and the gauge terms which are flavour blind. Before continuing with the IA, a relevant question is what role is played by the consistency relations \cite{Feruglio:2012cw, Holthausen:2012dk} in this type of analysis.

\begin{figure}
\centering
		\includegraphics[width=6.6 cm]{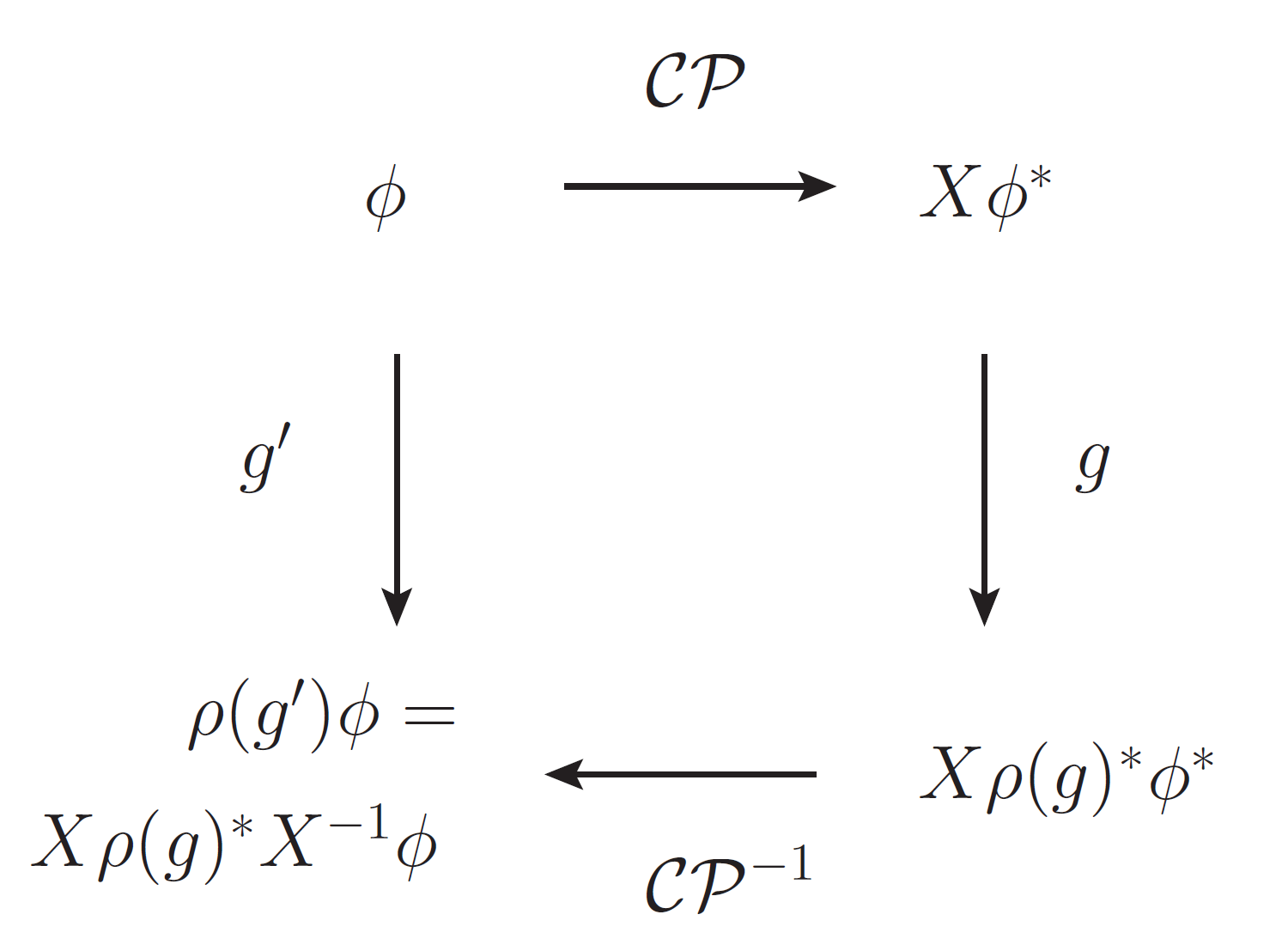}
\caption{The \CP transformation $X$ is consistent with the group $G$, as following it with the transformation $\rho(g)$ associated with element $g$ of $G$, and then with $X^{-1}$, is equivalent to the transformation $\rho(g')$ associated with some other element $g'$ of $G$. 		\label{fig:cons}}
\end{figure}

The consistency relations can be obtained by considering that a Lagrangian invariant under both a family symmetry and a \CP symmetry should be the same whether one considers doing a consistent \CP transformation before or after a family symmetry transformation. The concept is illustrated in Fig.~\ref{fig:cons} for a field $\phi$.
When this is considered rigorously
one obtains a relationship between \CP transformations $X$ and family symmetry transformations $\rho(g)$
\begin{equation}
\label{eq:consistency}
X \rho(g)^{*} X^{-1}=\rho(g'),
\qquad g' \in G\,.
\end{equation}
If we find an explicit \CP transformation that leaves the Lagrangian which respects some family symmetry $G$ invariant (even one is enough) then we can be quite sure that the theory conserves \CP as well as respecting the $G$. In this case the consistency relation in Eq.\ref{eq:consistency} is automatically satisfied. This is clear since under a \CP transformation, followed by a family symmetry transformation, followed by another \CP transformation, etc., leaves the Lagrangian invariant,
\begin{equation}
\mathcal{L} \stackrel{\mathcal{CP}}{\longrightarrow}  \mathcal{L}  \stackrel{G}{\longrightarrow}   \mathcal{L}
\stackrel{\mathcal{CP}}{\longrightarrow}  \mathcal{L} \stackrel{G}{\longrightarrow}  \mathcal{L}  \longrightarrow  \ldots
\end{equation}
The consistency of \CP and $G$ is clear since the Lagrangian is left invariant at each stage.
However we can be even more explicit than this in order to demonstrate the equivalence of the two approaches.

Consider a mass term $m$ in the Lagrangian, then define
\begin{equation}
H=mm^{\dagger} \,.
\label{H1}
\end{equation}
Suppose that the Lagrangian is invariant under some family symmetry transformation, 
$\rho(g)$, then this implies that the mass term
in the Lagrangian remains unchanged under a family symmetry transformation and hence
\begin{equation}
\rho(g)^{\dagger}H\rho(g)=H \,.
\label{cond00}
\end{equation}
The condition for the invariance of the Lagrangian under a \CP transformation, $X$, requires
that the mass term swaps with the $H.c.$ mass term hence,
\begin{equation}
X^{\dagger}HX=H^* \,.
\label{cond0}
\end{equation}
Taking the complex conjugate of Eq.\ref{cond00} we find,
\begin{equation}
(\rho(g)^{\dagger})^*H^*\rho(g)^*=H^*=X^{\dagger}HX \,,
\label{cond01}
\end{equation}
using Eq.\ref{cond0} for the last equality. 
From Eq.\ref{cond01} and Eq.\ref{cond0} we obtain,
\begin{equation}
(\rho(g)^{\dagger})^*X^{\dagger}HX\rho(g)^*=X^{\dagger}HX \,,
\label{cond02}
\end{equation}
hence
\begin{equation}
X(\rho(g)^{\dagger})^*X^{\dagger}HX\rho(g)^*X^{\dagger}=H=\rho(g')^{\dagger}H\rho(g') \,,
\label{cond03}
\end{equation}
where we have used Eq.\ref{cond00} but for a different group element $g'$ in the last equality.
By comparing both sides of Eq.\ref{cond03} we identify,
\begin{equation}
X\rho(g)^*X^{\dagger}=\rho(g') \,,
\end{equation}
which is just the consistency condition in Eq.\ref{eq:consistency}.

As we further illustrate in the following sections, using the IA one need not specify a \CP transformation. For a given Lagrangian it is sufficient to input the invariance conditions imposed by the symmetries. This makes the IA very useful to study \CP violation in the presence of family symmetries.

\section{$\Delta(27)$ group theory \label{D27}}

The group $\Delta(27)$ \cite{Branco:1983tn, deMedeirosVarzielas:2006fc,Ma:2006ip}, a member of the $\Delta(3 n^2)$ subgroups of $SU(3)$ \cite{Luhn:2007uq, Ishimori:2010au}, has featured prominently in studies related with \CP. It leads to geometrical \CP violation as shown in \cite{Branco:1983tn} and \cite{deMedeirosVarzielas:2011zw, Varzielas:2012nn, Bhattacharyya:2012pi, Varzielas:2013sla, Varzielas:2013eta}, a feature that was analysed using different methods by \cite{Holthausen:2012dk, Chen:2014tpa, Ivanov:2014doa, Fallbacher:2015rea} \footnote{Spontaneous geometrical \CP violation has also been found in other groups \cite{Varzielas:2012pd, Ivanov:2013nla,Varzielas:2013zbp}.}. In addition, the \CP transformations consistent with $\Delta(27)$ triplets were presented in \cite{Nishi:2013jqa}.

Using the IA, \cite{Branco:2015hea} found that a specific $\Delta(27)$ invariant Lagrangian features a different type of geometrical \CP violation, where \CP is explicitly violated (rather than spontaneously). In this paper we go beyond this result, considering many $\Delta(27)$ invariant Lagrangians to study in depth how the interplay between $\Delta(27)$ and \CP changes depending on the representations and how they couple with one another.

To do so, some understanding of the group properties is required. $\Delta(27)$ has three $Z_3$ generators \cite{Luhn:2007uq, Ishimori:2010au} but we only need to use two, which we refer to as $c$ (for cyclic, with $c^3=\Id$) and $d$ (for diagonal, with $d^3=\Id$). This notation refers to their respective 3-dimensional representation matrices in the basis we use.
We define $\omega \equiv e^{\ii 2 \pi/3}$.
Starting with the 9 distinct singlets which we conveniently label as $1_{ij}$, the generators are represented by $c_{1_{ij}}=\omega^i$ and $d_{1_{ij}}=\omega^j$  for that particular singlet. A field transforming as a $1_{00}$ (trivial singlet) is explicitly invariant under $\Delta(27)$ transformations, and the other 8 singlets simply get multiplied by the respective powers of $\omega$ when acted upon by $c$ or $d$.
The other irreducible representations of $\Delta(27)$ are triplets, two distinct ones which we take as $3_{01}$ and $3_{02}$. The generator $c$ is represented equally for both
\begin{equation}
c_{3_{ij}}=
\begin{pmatrix}
	0 & 1 & 0 \\
	0 & 0 & 1 \\
	1 & 0 & 0
\end{pmatrix} \,.
\end{equation}
$d$ is represented as a diagonal matrix with entries that are powers of $\omega$, with the exponents denoted by the indices of the triplet representation
\begin{equation}
d_{3_{ij}}=
\begin{pmatrix}
	\omega^i & 0 & 0 \\
	0 & \omega^j & 0 \\
	0 & 0 & \omega^{-i-j}
\end{pmatrix} \,.
\end{equation}
The determinant of the matrices is 1 ($\Delta(27)$ is a subgroup of $SU(3)$) and the two indices identify $d_{3_{01}}= \mathrm{diag} (1, \omega, \omega^2)$, $d_{3_{02}}= \mathrm{diag}(1, \omega^2, \omega)$. The representations $3_{01}$ and $3_{02}$ behave as a triplet and anti-triplet, so in analogy with $SU(3)$ we refer to them mostly as the $3$ and $\bar{3}$ representations. The subscript notation is useful to remember the powers of $\omega$ that each component transforms with under $d_{3_{ij}}$ so we refer to it occasionally throughout the paper, e.g. if we take $A=(a_1,a_2,a_3)_{01}$ transforming as triplet $3=3_{01}$ and $\bar{B}=(\bar{b}_1,\bar{b}_2,\bar{b}_3)_{02}$ transforming as (anti-)triplet $\bar{3}=3_{02}$, the explicit construction of the trivial singlet is $(A \bar{B})_{00}=(a_1 \bar{b}_1 + a_2 \bar{b}_2 + a_3 \bar{b}_3)_{00}$. This can be verified by acting on $A$ and $\bar{B}$ with generators $c$ and $d$ and checking that the prescribed $(A \bar{B})_{00}$ remains invariant.
Indeed, $3 \otimes \bar{3} = \sum_{i,j} 1_{ij}$ and rules for constructing the non-trivial singlets from triplet and anti-triplet follow
\begin{equation}
(a_2 \bar{b}_1 + a_3 \bar{b}_2 + a_1 \bar{b}_3)_{01} \,,
\label{AB01}
\end{equation}
\begin{equation}
(a_1 \bar{b}_2 + a_2 \bar{b}_3 + a_3 \bar{b}_1)_{02} \,,
\label{AB02}
\end{equation}
\begin{equation}
(a_1 \bar{b}_1 + \omega^2 a_2 \bar{b}_2 + \omega a_3 \bar{b}_3)_{10} \,,
\label{AB10}
\end{equation}
\begin{equation}
(a_2 \bar{b}_1 + \omega^2 a_3 \bar{b}_2 + \omega a_1 \bar{b}_3)_{11} \,,
\label{AB11}
\end{equation}
\begin{equation}
(\omega^2 a_1 \bar{b}_2 + \omega a_2 \bar{b}_3 + a_3 \bar{b}_1)_{12} \,,
\label{AB12}
\end{equation}
\begin{equation}
(a_1 \bar{b}_1 + \omega a_2 \bar{b}_2 + \omega^2 a_3 \bar{b}_3)_{20} \,,
\label{AB20}
\end{equation}
\begin{equation}
(a_2 \bar{b}_1 + \omega a_3 \bar{b}_2 + \omega^2 a_1 \bar{b}_3)_{21} \,,
\label{AB21}
\end{equation}
\begin{equation}
(\omega a_1 \bar{b}_2 + \omega^2 a_2 \bar{b}_3 + a_3 \bar{b}_1)_{22} \,.
\label{AB22}
\end{equation}
All these can be verified by acting on the triplets with the generators and tracking how each product transforms.

\section{Just singlets \label{singlets}}

To illustrate how the IA would proceed, we start by considering Yukawa-like terms without $\Delta(27)$ triplets. Throughout we refer to fields transforming as singlets under $\Delta(27)$ as $h_{ij}$ where the subscript refers to the field being assigned as a $1_{ij}$ under $\Delta(27)$.

A simple example where the field content is $h_{00}$, $h_{01}$, $h_{10}$ would have Yukawa-like terms
\begin{align}
\mathcal{L}_{III} =  & z_{00} h_{00} h_{00} h_{00} + z_{01} h_{01} h_{01} h_{01} + z_{10} h_{10} h_{10} h_{10} \notag \\
+ & y_{00} h_{00} h_{00} h_{00}^\dagger + y_{01} h_{00} h_{01} h_{01}^\dagger + y_{10} h_{00} h_{01} h_{01}^\dagger + H.c. \,.
\label{Ls}
\end{align}
It is clear there are further $\Delta(27)$ invariant terms such as $h_{00} h_{00}$, but for the sake of illustrating the IA we consider the \CP properties of the Yukawa-like terms in $\mathcal{L}_{III}$ by itself as the only part of the full Lagrangian that can violate \CP (i.e. as the $\mathcal{L}_{rem.}$ for this case). \footnote{We are limiting ourselves to tri-linear terms even for field contents only with scalars, because we are keeping in mind renormalisable Yukawa terms.}
The next step in the IA is to consider the most general \CP transformation for each field consistent with the kinetic terms etc., in this case this means each singlet transforms with its own phase which we denote as $p_{ij}$
\begin{align}
h_{ij} \rightarrow e^{\ii p_{ij}} h_{ij}^*  \,.
\label{hij}
\end{align}
When we apply these transformations on $\mathcal{L}_{III}$ and demand it remains invariant, we obtain a set of necessary and sufficient conditions for \CP conservation restricting the parameters in $\mathcal{L}_{III}$
\begin{align}
z_{00} e^{\ii 3 p_{00}}= z_{00}^* \,, \\
z_{01} e^{\ii 3 p_{01}}= z_{01}^* \,, \\
z_{10} e^{\ii 3 p_{10}}= z_{10}^* \,, \\
y_{00} e^{\ii p_{00}}= y_{00}^* \, \\
y_{01} e^{\ii p_{00}}= y_{01}^* \,, \\
y_{10} e^{\ii p_{00}}= y_{10}^* \,.
\label{ys}
\end{align}
We note that the \CP conservation conditions on couplings $y_{01}$ and $y_{10}$ are independent of $p_{01}$, $p_{10}$.
To build CPIs we combine conditions that cancel dependence on the \CP transformations, which involving  $y_{01}$ and $y_{10}$ requires only the cancellation of all dependence on $p_{00}$. A simple and useful CPI is then $\mathrm{Im} [ y_{01} y_{10}^*]$, as
\begin{equation}
y_{01} y_{10}^\dagger = (y_{01} y_{10}^\dagger)^* \to
\mathrm{Im} [ y_{01} y_{10}^*] = 0 \,,
\label{0110_rel}
\end{equation}
where the $y_{ij}$ are complex numbers so $y_{ij}^\dagger = y_{ij}^*$.
The CPI vanishing is a necessary (but not necessarily sufficient) condition for \CP conservation and it constrains the relative phase between the two couplings.

There are also CPIs of this type constraining the relative phases between $y_{00}$ and the other two $y_{ij}$ couplings. CPIs involving $z_{00}$ can also be built noting that the CPI needs to cube the other couplings e.g. $\mathrm{Im} [ z_{00}^* y_{ij}^3]$, as
\begin{equation}
z_{00}^\dagger y_{ij}^3 = (z_{00}^\dagger y_{ij}^3)^* \to
\mathrm{Im} [ z_{00}^* y_{ij}^3]=0 \,.
\end{equation}
If any of these CPI are non-zero, then \CP is violated.
Other CPIs like $\mathrm{Im} [z_{ij} z_{ij}^*]$ (or $\mathrm{Im} [y_{ij} y_{ij}^*]$) do not provide useful constraints as they automatically vanish.

As a first generalisation of $\mathcal{L}_{III}$ we add a field $h_{02}$. This allows a mixed invariant whose coupling constant will be sensitive to the \CP transformation of the non-trivial singlets. We continue to consider only Yukawa-like terms in $\mathcal{L}_{rem.}$ so we write
\begin{equation}
\mathcal{L}_{IV} = \mathcal{L}_{III} + (z_{02} h_{02} h_{02} h_{02} + y_{02} h_{00} h_{02} h_{02}^\dagger + y_{1} h_{00} h_{01} h_{02} + H.c.) \,.
\label{Ls2}
\end{equation}
The new \CP conditions are 
\begin{align}
z_{02} e^{\ii 3 p_{02}}= z_{02}^* \,, \\
y_{02} e^{\ii p_{00}}= y_{02}^* \,, \\
y_{1} e^{\ii (p_{00}+p_{01}+p_{02})}= y_{1}^* \,.
\label{ys2}
\end{align}
With the added couplings, we have from the direct generalisations of Eq.~\ref{0110_rel} CPIs  involving $y_{02}$ like
\begin{align}
\mathrm{Im} &[y_{00} y_{02}^*] \,, \\
\mathrm{Im} &[y_{02} y_{01}^*] \,, \\
\mathrm{Im} &[y_{01} y_{02}^*] \,,
\end{align}
which constrain the relative phases of the $y_{ij}$ couplings. It is however not possible to obtain such simple, useful constraints on the relative phase of the mixed coupling $y_{1}$, because it involves $p_{01}$ and $p_{02}$. The simplest CPI, $\mathrm{Im} [y_{1} y_{1}*]$
automatically vanishes and is therefore not useful. The conclusion from this CPI alone would be that the phase of $y_{1}$ is unconstrained by requiring $\mathcal{L}_{IV}$ to conserve \CP.
However, we note that by using a more complicated invariant involving $z_{01}$, $z_{02}$ and either $z_{00}$ or 3 insertions of any $y_{ij}$, we can build non-trivial CPIs involving $y_{1}$, such as
\begin{align}
\mathrm{Im} [y_{1}^3 z_{00}^* z_{01}^* z_{02}^*] \,, \label{z_rel} \\
\mathrm{Im} [y_{1}^3 y_{ij}^{* 3} z_{01}^* z_{02}^*] \,.
\end{align}

The situation involving the coupling of mixed terms of the type $y_{1}$ qualitatively changes if there are sufficient mixed terms.
Consider now a field content with all 9 $\Delta(27)$ singlets $h_{ij}$. To reduce the number of allowed terms we may impose a $Z_3$ symmetry where each $h_{ij}$ transforms equally. Such a $Z_{3}$ symmetry forces $y_{ij}=0$ (in addition it forbids a multitude of other Yukawa-like terms like $h_{01} h_{10} h_{11}^\dagger$, where $h_{11}^\dagger$ would play the role of $h_{22}$).
There are 9 Yukawa-like terms like $z_{00} h_{00} h_{00} h_{00}$ (one for each singlet). CPIs involving the $z_{ij}$ will be like Eq.~\ref{z_rel}.
Focusing solely on the mixed terms like $y_{1} h_{00} h_{01} h_{02}$, there are 12 combinations of singlets whose indices add up to a mixed invariant
\begin{align}
\mathcal{L}_{IX} =  y_{1} h_{00} h_{01} h_{02} + y_{2} h_{00} h_{10} h_{20} + y_{3} h_{00} h_{11} h_{22} + y_{4} h_{00} h_{12} h_{21}+& \notag \\
 y_{5} h_{01} h_{10} h_{22} + y_{6} h_{01} h_{11} h_{21} + y_{7} h_{01} h_{12} h_{20} +& \notag \\
 y_{8} h_{02} h_{10} h_{21} + y_{9} h_{02} h_{11} h_{20} + y_{10} h_{02} h_{12} h_{22} +& \notag \\
 y_{11} h_{10} h_{11} h_{12} + y_{12} h_{20} h_{21} h_{22} +& H.c. \,. \label{Ls9}
\end{align}
The \CP conservation condition for each coupling depends on the 3 phases of the respective singlets, e.g.
\begin{align}
y_{1} e^{\ii (p_{00}+p_{01}+p_{02})} = y_{1}^* \,, \\
y_{2} e^{\ii (p_{00}+p_{10}+p_{20})} = y_{2}^* \,, \\
y_{6} e^{\ii (p_{01}+p_{11}+p_{21})} = y_{6}^* \,, \\
y_{10} e^{\ii (p_{02}+p_{12}+p_{22})}= y_{10}^* \,, \\
y_{11} e^{\ii (p_{10}+p_{11}+p_{12})}= y_{11}^* \,, \\
y_{12} e^{\ii (p_{20}+p_{21}+p_{22})}= y_{12}^* \,.
\label{ys9}
\end{align}
With these conditions it is now possible to combine several of the mixed couplings to form a CPI. An example is
\begin{equation}
\mathrm{Im} [y_1 y_2^* y_6^* y_{10}^* y_{11} y_{12}] \,,
\end{equation}
meaning this particular combination of couplings is constrained by \CP conservation to be real. Other combinations of this type can be built from the couplings in $\mathcal{L}_{IX}$.

\section{One triplet}
\label{one}

We will now consider Lagrangians involving Yukawa-like couplings with just one $\Delta(27)$ triplet \cite{Varzielas:2015fxa}. In order to make invariants, the terms will necessarily involve the conjugate of that triplet.
In this case it is not possible to construct Yukawa couplings involving a $\Delta(27)$ triplet Weyl fermion. If $F \sim 3$, while we do have $F^\dagger \sim \bar{3}$, the $\Delta(27)$ invariant of type $(F F^\dagger)_{ij} h_{kl}$ is not itself invariant under Lorentz symmetry as e.g. $F \sim (1/2,0)$ implies $F^\dagger \sim (0, 1/2)$.
We construct the Lagrangian with one scalar triplet $\phi \sim 3$ and 2 scalar singlets $h_{01}$, $h_{10}$, such that the Yukawa-like terms are
\begin{equation}
\mathcal{L}_{2s} = y_{01} (\phi \phi^*)_{02} h_{01} + y_{10} (\phi \phi^*)_{20} h_{10} + H.c. \,.
\label{L2s}
\end{equation}
The most general \CP transformations are
\begin{align}
h_{01} &\rightarrow e^{\ii p_{01}} h_{01}^* \,, \\
h_{10} &\rightarrow e^{\ii p_{10}} h_{10}^* \,, \\
\phi &\rightarrow U^{*} \phi^* \,,
\label{CP2s}
\end{align}
where $U$ is a general unitary matrix.
Assuming \CP invariance of $\mathcal{L}_{2s}$ and using matrices $Y_{ij}$ corresponding to the couplings $y_{ij}$ we have
\begin{align}
U^\dagger Y_{01} U e^{\ii p_{01}}= Y_{01}^* \,, \\
U^\dagger Y_{10} U e^{\ii p_{10}}= Y_{10}^* \,.
\label{Y2s}
\end{align}
We can build a useful CPI for this Lagrangian \cite{Varzielas:2015fxa}
\begin{equation}
I_{2s} \equiv \mathrm{Im} \Tr (Y_{01} Y_{10}^\dagger Y_{01}^\dagger Y_{10}) \,.
\label{I2s}
\end{equation}
The CPI applies for any matrices $Y_{01}$ and $Y_{10}$. Imposing $\Delta(27)$ invariance we have from Eq.~\ref{AB01},\ref{AB10}
\begin{equation}
Y_{01}=y_{01}
\begin{pmatrix}
0 & 1 & 0\\
0 & 0 & 1\\
1 & 0 & 0
\end{pmatrix} \,,
\end{equation}
\begin{equation}
Y_{10}=y_{10}
\begin{pmatrix}
1 & 0 & 0\\
0 & \omega & 0\\
0 & 0 & \omega^2
\end{pmatrix} \,,
\end{equation}
which we input into $I_{2s}$ and obtain
\begin{equation}
I_{2s} = \mathrm{Im} (3 \omega^2 |y_{01}|^2 |y_{10}|^2) \,.
\end{equation}
Finding a non-vanishing CPI means that \CP is violated, which clearly happens for any non-zero values of $y_{01}$ and $y_{10}$. Given that \CP is explicitly violated by a phase only originating from the group structure and not from arbitrary Lagrangian parameters, this is a minimal case with explicit geometrical \CP violation \cite{Branco:2015hea,Varzielas:2015fxa}.

In \cite{Chen:2014tpa} it was pointed out that $\Delta(27)$ provides an example of a group where not all Clebsch-Gordan coefficients can be made real by a change of basis, when several of the singlets are used. Indeed, this fact was already referred in the earlier $\Delta(27)$ works \cite{Bhattacharyya:2012pi,Varzielas:2013sla,Varzielas:2013eta} where CP is violated spontaneously and therefore only a few singlets were used. The change of basis analysis presented explicitly in \cite{Varzielas:2015fxa} further clarifies the connection between the inevitability of complex Clebsch-Gordan coefficients (which are basis-dependent) and the presence of multiple singlets. The physical consequences are of course basis-independent as illustrated elegantly in the invariant approach, and depend crucially on the field content, not just of singlets but also of triplets as shown in the following sections.

\section{Two triplets \label{2triplets}}

We continue our exploration of unbroken $\Delta(27)$ invariant Lagrangians with Yukawa-like terms by considering in detail the class with two distinct 3-dimensional representations. We take these to be explicitly a triplet $Q \sim 3$ and an anti-triplet $d^c \sim \bar{3}$. \footnote{This is mostly equivalent to introducing another triplet $d$, whose $H.c.$ $d^\dagger$ would transform as an anti-triplet.}

The notation we are following is suggestive of identifying the $3$ and $\bar{3}$ as fermions, as considered in \cite{Branco:2015hea}. Nevertheless, it is also possible to consider just scalars \cite{Varzielas:2015fxa}, or that one of the triplets is a scalar and Yukawa couplings are formed by having $\Delta(27)$ singlet fermions, as considered in \cite{deMedeirosVarzielas:2011zw, Varzielas:2012nn, Bhattacharyya:2012pi, Varzielas:2013sla, Varzielas:2013eta}. The conclusions we derive with the IA apply to all cases.

As a further point of notation, if e.g. $Q$ does refer to the SM quark fields, the most general \CP transformation consists in
\begin{equation}
(\mathcal{CP}) Q (\mathcal{CP})^\dagger =i U \gamma^0 \mathcal{C} \bar{Q}^T \,.
\label{fermionCP}
\end{equation}
Here we need not specify if $Q$ is a fermion or scalar, and in particular we are more concerned with identifying which matrix corresponds to each field (in Eq.~\ref{fermionCP}, $U$ corresponds to $Q$). For these reasons we use a simplified notation along the lines of
\begin{equation}
Q \rightarrow \mathcal{CP} Q = U_Q^T Q^* \,,
\label{scalarCP}
\end{equation}
which is exact for scalars, and is more convenient to identify which general \CP transformation corresponds to each field (in Eq.~\ref{scalarCP}, $U_Q^T$ corresponds to $Q$).

We assume the $\mathcal{L}_{rem.}$ part of the Lagrangian consists of Yukawa-like terms between triplet $Q \sim 3$, anti-triplet $d^c \sim \bar{3}$, and singlets $h_{ij} \sim 1_{ij}$. 
\footnote{Indeed it is relatively straightforward to forbid additional tri-linears if these are actually Yukawa terms between fermions and scalars.}

This class of Lagrangians is a good framework to illustrate several interesting points, so we go into some detail of what happens when varying the number of coupled singlets. 
The first model we consider of this type is with $Q$, $d^c$ and 2 singlets $h_{10}$ and $h_{01}$ \cite{Varzielas:2015fxa}
\begin{equation}
\mathcal{L}_3 =y_{10} (Q d^c)_{20} h_{10} + y_{01} (Q d^c)_{02} h_{01} + H.c.
\label{L3}
\end{equation}

\subsection{Adding a specific \CP symmetry}

Before applying the IA to Lagrangians of this type, lets consider what happens when applying a specific \CP transformation.

Arguably the simplest \CP transformation is the trivial \CP transformation, which we refer to as $\mathcal{CP}_1$. This corresponds to $U_Q = \Id$ in Eq.~\ref{scalarCP}, i.e.
\begin{equation}
Q \rightarrow \mathcal{CP}_1 Q = Q^* = (Q_1^*, Q_2^*, Q_3^*)_{02} \,,
\label{CP1A}
\end{equation}
where we used the subscript to denote that $Q^*$ transforms as a $\bar{3} = 3_{02}$, given that under action by generator $d$, $Q_2 \rightarrow \omega Q_2$ and therefore we must have $Q_2^* \rightarrow \omega^2 Q_2^*$ under $d$ (as expected from complex conjugation). Similarly, 
\begin{equation}
\mathcal{CP}_1 d^c = d^{c *} = (d_1^{c *},d_2^{c *},d_3^{c *})_{01} \,.
\label{CP1B}
\end{equation}
For the two $\Delta(27)$ singlets $h_{10}$ and $h_{01}$ in $\mathcal{L}_3$
\begin{align}
\mathcal{CP}_1 h_{10} =& h_{10}^* \,, \label{CP1C} \\
\mathcal{CP}_1 h_{01} =& h_{01}^* \,, \label{CP1D}
\end{align}
where the conjugated versions transform under $\Delta(27)$ as $1_{20}$ and $1_{02}$ respectively.
This means that under the trivial \CP transformation all four fields go into their respective conjugate $\Delta(27)$ representations.

The Yukawa-like terms in $\mathcal{L}_3$ are explicitly invariant under $\Delta(27)$ and $y_{10}$, $y_{01}$ are arbitrary complex numbers. We now impose additionally that $\mathcal{L}_3$ is invariant under $\mathcal{CP}_1$. For the $y_{10}$ coupling, expanding from Eq.~\ref{AB20} and using Eq.~\ref{CP1A}, \ref{CP1B}, \ref{CP1C}
\begin{equation}
 y_{10} (Q_1 d_1^{c} + \omega Q_2 d_2^{c} + \omega^2 Q_3 d_3^{c})_{20} h_{10} \rightarrow y_{10} (Q_1^* d_1^{c *} + \omega Q_2^* d_2^{c *} + \omega^2 Q_3^* d_3^{c *})_{20} h_{10}^* \,.
\label{CP1yc}
\end{equation}
In identifying how $(Q d^c)_{20}$ has transformed under $\mathcal{CP}_1$, note the  $\mathcal{CP}_1$-transformed product $(Q d^c)_{20}$ still transforms as a $1_{20}$ under $\Delta(27)$, as it picks up a phase of $\omega^2$ when acted by $c$.
For the $y_{01}$ coupling, expanding from Eq.~\ref{AB02} and using Eq.~\ref{CP1D}
\begin{equation}
y_{01} (Q_1 d_2^{c} + Q_2 d_3^{c} + Q_3 d_1^{c})_{02} h_{01} \rightarrow y_{01} (Q_1^* d_2^{c *} + Q_2^* d_3^{c *} + Q_3^* d_1^{c *})_{01} h_{01}^* \,.
\label{CP1yd}
\end{equation}
In contrast, because of the action in Eq.~\ref{CP1B}, $d_2^{c *}$ picks up a phase $\omega$ when acted by $d$, we identify that the  $\mathcal{CP}_1$-transformed product $(Q d^c)_{02}$ transforms as a $1_{01}$ under $\Delta(27)$.
What are then the physical consequences of imposing $\mathcal{CP}_1$ on the $\Delta(27)$-invariant $\mathcal{L}_3$?
We need to compare Eq.~\ref{CP1yc} and Eq.~\ref{CP1yd} to the $H.c.$ part of $\mathcal{L}_3$. In the case of Eq.~\ref{CP1yd} this reveals exactly the same expression, except that $y_{01}^*$ appears, therefore the conclusion is clear - imposing $\mathcal{CP}_1$ on $\mathcal{L}_3$ forces $y_{01}^*=y_{01}$.
However, when we compare what we obtained in Eq.~\ref{CP1yc} to 
\begin{equation}
y_{10}^* (Q_1^* d_1^{c *} + \omega^2 Q_2^* d_2^{c *} + \omega Q_3^* d_3^{c *})_{10} h_{10}^* \,,
\end{equation}
the only way to make the expressions match (to have $\mathcal{L}_3$ be invariant under$\mathcal{CP}_1$) is to require $y_{10}^*=y_{10}=0$. That there is some incompatibility with $y_{10}$ was already hinted at by the fact that Eq.~\ref{CP1yc} is explicitly not invariant under $\Delta(27)$, which we denoted through the subscripts - $(...)_{20}$ transforms as a $1_{20}$, as does $h_{10}^*$.

Our interpretation of these results is not that $\mathcal{CP}_1$ becomes incompatible with $\Delta(27)$ when the theory includes field $h_{10}$ together with $Q$ and $d^c$. Rather, that it is always possible to add $\mathcal{CP}_1$ to a $\Delta(27)$ invariant Lagrangian regardless of the field content, but we interpret this result as indicating that there will be physical consequences on the couplings that make the theory consistent by making the Lagrangian invariant under the full symmetry imposed. In other words, it is more correct to state the incompatibility is not with the field content but rather with the couplings.

That $\mathcal{CP}_1$ restricts $\mathcal{L}_3$ does not mean that \CP is violated though, and in fact a \CP transformation can be defined that leaves the Lagrangian invariant. A good way to check $\mathcal{L}_3$ is \CP conserving is to use the IA. Before doing so, lets examine the effects of another possible \CP transformation $\mathcal{CP}_2$, keeping the same transformations as $\mathcal{CP}_1$ for the singlets but where the triplet transforms as 
\begin{equation}
Q \rightarrow \mathcal{CP}_2 Q = Q'^* = (Q_1^*, Q_3^*, Q_2^*)_{01} \,,
\label{CP2A}
\end{equation}
where some components swapped positions. Similarly, 
\begin{equation}
\mathcal{CP}_2 d^c = d'^{c * } = (d_1^{c *}, d_3^{c *}, d_2^{c *})_{02} \,.
\label{CP2B}
\end{equation}
Note that unlike what happens with $\mathcal{CP}_1$, given that under action from generator $d$ we have $Q_3 \rightarrow \omega^2 Q_3$ and therefore $Q_3^* \rightarrow \omega Q_3^*$, this $Q'^*$ transforms as a $3 = 3_{01}$, and similarly this $d'^{c *}$ transform as a $\bar{3}=3_{02}$.

Checking how the $\mathcal{L}_{3}$ terms transform under $\mathcal{CP}_1$ we have 
\begin{equation}
y_{10} (Q_1 d_1^{c} + \omega Q_2 d_2^{c} + \omega^2 Q_3 d_3^{c})_{20} h_{10} \rightarrow y_{10} (Q_1^* d_1^{c *} + \omega Q_3^* d_3^{c *} + \omega^2 Q_2^* d_2^{c *})_{10} h_{10}^* \,.
\label{CP2yc}
\end{equation}
In identifying how $(Q d^c)_{20}$ has transformed under $\mathcal{CP}_2$, it now picks up a phase of $\omega$ when acted by $c$.
The other combination
\begin{equation}
y_{01} (Q_1 d_2^{c} + Q_2 d_3^{c} + Q_3 d_1^{c})_{02} h_{01} \rightarrow y_{01} (Q_1^* d_3^{c *} + Q_3^* d_2^{c *} + Q_2^* d_1^{c *})_{02} h_{01}^* \,,
\label{CP2yd}
\end{equation}
and one can verify that the  $(Q d^c)_{20}$ has transformed under $\mathcal{CP}_2$ into an expression that picks up a phase of $\omega^2$ when acted by $d$.
By comparing Eq.~\ref{CP2yc}, \ref{CP2yd} we conclude that by imposing $\mathcal{CP}_2$ invariance on Eq.~\ref{L3} we are forcing $y_{10}$ to be real and $y_{01}=0$ (contrast with $\mathcal{CP}_1$).

While we used $\Delta(27)$ and $\mathcal{CP}_1$ (and $\mathcal{CP}_2$) in this explicit example, our interpretation is general - we have the freedom to impose any family symmetry (discrete or not) together with any \CP symmetry. Eventually what may happen in extreme cases, is that it will not be possible to form non-trivial combinations that are invariant under both symmetries. We feel it is important to stress that in this interpretation, the \CP symmetry is  consistently treated equally to other symmetries - the transformation is defined and it has consequences for the Lagrangian.

It is important to stress again that even if imposing a specific \CP symmetry on a theory restricts the couplings of the Lagrangian, this does not mean that the Lagrangian violates \CP. $\mathcal{L}_3$ is an example of that as we now show using the IA. It is convenient to first rewrite $\mathcal{L}_3$ in terms of coupling matrices $Y_{01}$ and $Y_{10}$
\begin{equation}
\mathcal{L}_3 =Q Y_{10} d^c h_{10} + Q Y_{01} d^c h_{01} + H.c. \,,
\label{L3Y}
\end{equation}
and to specify the general \CP transformation properties
\begin{align}
h_{01} &\rightarrow e^{\ii p_{01}} h_{01}^* \,, \notag \\
h_{10} &\rightarrow e^{\ii p_{10}} h_{10}^* \,, \notag \\
Q &\rightarrow U_Q^T Q^* \,, \notag \\
d^c &\rightarrow U_d d^{c *}\,.
\label{L3CP}
\end{align}
One can verify that $\mathcal{CP}_1$ and $\mathcal{CP}_2$ are particular cases of this general \CP transformation.
Imposing invariance under the general \CP transformation requires
\begin{align}
U_Q Y_{01} U_d e^{\ii p_{01}}= Y_{01}^* \,, \\
U_Q Y_{10} U_d e^{\ii p_{10}}= Y_{10}^* \,.
\label{Y3sQ}
\end{align}
We again wish to build combinations of the Yukawa-like couplings that eliminate the dependence on the general transformation. Unlike what we considered with $\mathcal{CP}_1$ and $\mathcal{CP}_2$, $Q$ and $d^c$ transform in general with distinct unrelated matrices. This forces CPIs to alternate between $Y_{ij}$ and $Y_{kl}^\dagger$. It is therefore convenient to define the Hermitian combinations $G_{ij} \equiv Y_{ij}^\dagger Y_{ij}$ and $H_{ij}=Y_{ij} Y_{ij}^\dagger$ (similar Hermitian combinations appear in the SM CPI, $\Tr [H_u, H_d]^3$).
They are useful because general \CP invariance requires
\begin{align}
U_d^\dagger G_{ij} U_d = G_{ij}^* \,, \\
U_Q H_{ij} U_Q^\dagger = H_{ij}^* \,,
\label{GH3sQ}
\end{align}
involving for each only the matrix associated with $d^c$ and $Q$, respectively.

The CPIs are of the type (with integers $n_i$)
\begin{align}
\mathrm{Im} \Tr [ G_{01}^{n_1} G_{10}^{n_2} G_{01}^{n_3} (...)] \,,
\label{IA2sGp} \\
\mathrm{Im} \Tr [ H_{01}^{n_1} H_{10}^{n_2} H_{01}^{n_3} (...)] \,.
\label{IA2sHp}
\end{align}
When referring to CPIs of this type we are including also CPIs like
\begin{align}
 \mathrm{Im} \Tr [G_{01}^{n_1} Y_{01}^\dagger H_{10}^{n_2} Y_{01} G_{01}^{n_3} G_{10}^{n_4} (...)] \,, \label{splitG} \\
 \mathrm{Im} \Tr [ G_{01}^{n_1}  Y_{10}^\dagger Y_{01} G_{10}^{n_2} Y_{01}^\dagger Y_{10} G_{01}^{n_3} (...)] \label{splitG2} \,.
\end{align}
The first one alternates between $G_{01}$ and $H_{10}$ by having a single additional $Y_{01}^\dagger$ in the middle, and eventually goes back to $G_{01}$ due to the lone $Y_{01}$ that is required to cancel the dependence on $p_{01}$.
The second one is similar by having $Y_{10}^\dagger Y_{01}$ between $G_{01}$, then requiring $Y_{01}^\dagger Y_{10}$ somewhere else in order to cancel the dependence on $p_{01}$, $p_{10}$ - note the ordering of the inserted $Y_{ij}$ and $Y_{kl}^\dagger$ is not arbitrary.
It is also possible to mix and match odd and even insertions of $Y_{ij}$ and $Y_{kl}^\dagger$
\begin{equation}
\mathrm{Im} \Tr [ G_{01}^{n_1}  Y_{10}^\dagger Y_{01} G_{10}^{n_2} Y_{01}^\dagger  H_{01}^{n_3} Y_{10} G_{10}^{n_4} (...)] \label{splitG3} \,.
\end{equation}
We refer to the more complicated CPIs as being of the type Eq.~\ref{IA2sGp}, \ref{IA2sHp} because they are obtained by iteratively inserting some $G_{ij}$ or $H_{ij}$ inside an existing $G_{kl}$ or $H_{kl}$, thus separating the constituent $Y_{kl}$ and $Y_{kl}^\dagger$. \footnote{E.g. start with Eq.~\ref{IA2sGp} and split one of the $G_{01}$ in the middle with $H_{10}^{n_2}$ and redefine the integers to obtain Eq.~\ref{splitG}. Or in reverse, remove $G_{10}^{n_2}$ from Eq.~\ref{splitG3}, which allows an $H_{01}$ to be created and included into $H_{01}^{n_3}$ by redefining the integers and arrive back at Eq.~\ref{splitG} - which can then be related with Eq.~\ref{IA2sGp}.}

When considering the Yukawa-like matrices that comply with $\Delta(27)$ invariance
\begin{equation}
Y_{01}=y_{01}
\begin{pmatrix}
0 & 1 & 0\\
0 & 0 & 1\\
1 & 0 & 0
\end{pmatrix} \,,
\end{equation}
\begin{equation}
Y_{10}=y_{10}
\begin{pmatrix}
1 & 0 & 0\\
0 & \omega & 0\\
0 & 0 & \omega^2
\end{pmatrix} \,,
\end{equation}
that we have seen before, we have
\begin{equation}
G_{01} = H_{01} = G_{10}=H_{10}= \Id \,,
\end{equation}
and it is possible to check that all these CPIs vanish automatically, for any complex $y_{10}$, $y_{01}$. Therefore $\mathcal{L}_3$ conserves \CP for any $y_{10}$, $y_{01}$, even though it is not in general invariant under some of the particular \CP transformations like $\mathcal{CP}_1$ and $\mathcal{CP}_2$.

\subsection{Two singlets \label{sec:IAL3}}

The conclusion is a bit more general and applies beyond the pair of singlets $h_{01}$ and $h_{10}$. Using the IA it is relatively easy to verify that a Lagrangian of type $\mathcal{L}_3$ automatically conserves \CP for any two singlets $\Delta(27)$ (even if a specific \CP like $\mathcal{CP}_1$ imposes restrictions on these Lagrangians).
The field content is then $Q$, $d^c$ and any two singlets $h_{ij}$, $h_{kl}$. There are only two Yukawa-like terms and associated matrices
\begin{equation}
\mathcal{L}'_{3} = Q Y_{ij} d^c h_{ij} + Q Y_{kl} d^c h_{kl} + H.c. \,.
\end{equation}
The CPIs are trivial generalisations of the previous cases, e.g.
\begin{align}
\mathrm{Im}& \Tr [ G_{ij}^{n_1} G_{kl}^{n_2} G_{ij}^{n_3} (...)] \,,
\label{IA2sG} \\
\mathrm{Im}& \Tr [ H_{ij}^{n_1} H_{kl}^{n_2} H_{ij}^{n_3} (...)] \,,
\label{IA2sH}
\end{align}
and the other types discussed above. 
Due to invariance under $\Delta(27)$ we have
\begin{equation}
G_{ij} = H_{ij} = G_{kl}=H_{kl}= \Id \,,
\end{equation}
regardless of the singlets used. Using these relations allows one to conclude that all of these CPIs automatically vanish for any couplings $y_{ij}$, $y_{kl}$, meaning \CP is automatically conserved in Lagrangians of $\mathcal{L}_3$ type - Yukawa-like couplings of any 2 $\Delta(27)$ singlets to 2 triplets.
However, the possibility for explicit \CP violation exists for 3 singlets and beyond - see also
\cite{Chen:2014tpa, Fallbacher:2015rea}, but note that the conclusion depends on the number of triplets as well (e.g. 2 singlets are sufficient to violate \CP in the presence of only one triplet \cite{Varzielas:2015fxa}).

\subsection{Three singlets}

Applying the IA to Lagrangians of $\mathcal{L}_3$ type with 3 singlets allows us to identify explicit geometrical \CP violation. In this case we change notation from triplet $Q$ and anti-triplet $d^c$ to the notation used in the model of \cite{Branco:2015hea}, to allow an easier comparison. We introduce the SM fermions $L \sim 3$ and also $\nu^c \sim \bar{3}$ and singlet scalars $h_{00}$, $h_{01}$, $h_{10}$.
This model is a model of leptons, with a charged lepton Lagrangian that gives in this basis a diagonal mass matrix, with a $\Delta(27)$ triplet scalar $\phi \sim \bar{3}$
\cite{Branco:2015hea}
\begin{equation}
-y_e (L \phi)_{00} \,e^c_{00} -  y_{\mu}(L \phi)_{01} \,\mu^c_{02} -y_{\tau}(L \phi)_{02} \,\tau^c_{01}  + H.c. \,.
\label{D27cl}
\end{equation}
We focus here on the \CP properties of the neutrino Lagrangian $\mathcal{L}_{3s}$
\begin{equation}
\mathcal{L}_{3s} = y_{00} (L \nu^c)_{00} h_{00} + y_{01} (L \nu^c)_{02} h_{01} + y_{10} (L \nu^c)_{20} h_{10} + H.c. \,.
\label{L3s}
\end{equation}
This is similar to the $\mathcal{L}_3$ Lagrangian, with Yukawa-like terms between the triplets and the singlets.
The general \CP transformations are
\begin{align}
h_{00} &\rightarrow e^{\ii p_{00}} h_{00}^* \,, \\
h_{01} &\rightarrow e^{\ii p_{01}} h_{01}^* \,, \\
h_{10} &\rightarrow e^{\ii p_{10}} h_{10}^* \,, \\
L &\rightarrow U_L^T L^*\,, \\
\nu^c &\rightarrow U_\nu \nu^{c *}\,.
\label{CP3s}
\end{align}
Assuming \CP invariance of $\mathcal{L}_{3s}$ and expressing the couplings in terms of Yukawa matrices $Y_{ij}$, the \CP conservation requirements are
\begin{align}
U_L Y_{00} U_\nu e^{\ii p_{00}}= Y_{00}^* \,, \\
U_L Y_{01} U_\nu e^{\ii p_{01}}= Y_{01}^* \,, \\
U_L Y_{10} U_\nu e^{\ii p_{10}}= Y_{10}^* \,.
\label{Y3s}
\end{align}
$\Delta(27)$ imposes
\begin{equation}
Y_{00}=y_{00}
\begin{pmatrix}
1 & 0 & 0\\
0 & 1 & 0\\
0 & 0 & 1
\end{pmatrix} \,,
\end{equation}
\begin{equation}
Y_{01}=y_{01}
\begin{pmatrix}
0 & 1 & 0\\
0 & 0 & 1\\
1 & 0 & 0
\end{pmatrix} \,,
\end{equation}
\begin{equation}
Y_{10}=y_{10}
\begin{pmatrix}
1 & 0 & 0\\
0 & \omega & 0\\
0 & 0 & \omega^2
\end{pmatrix} \,.
\end{equation}
Because we have 3 Yukawa matrices, we can build a CPI that does not involve $G_{ij}$ or $H_{ij}$ combinations
\begin{equation}
I_{3s} \equiv \mathrm{Im} \Tr (Y_{00} Y_{01}^\dagger Y_{10} Y_{00}^\dagger Y_{01} Y_{10}^\dagger) \,.
\label{I3s}
\end{equation}
This CPI is qualitatively different from the ones that could be built with only 2 matrices. Indeed if we calculate it for this particular choice of 3 singlets
\begin{equation}
I_{3s}= \mathrm{Im} (3 \omega^2 |y_{00}|^2 |y_{01}|^2 |y_{10}|^2) \,.
\end{equation}
This means that in general, for arbitrary (non-zero) couplings, this Lagrangian violates \CP as the condition $I_{3s}=0$, necessary for \CP conservation, is not fulfilled. 
This is the case originally identified to have explicit geometrical \CP violation in \cite{Branco:2015hea}, with the phase appearing in $I_{3s}$ independent of the arbitrary phases of couplings.

We can again check what are the consequences of adding specific \CP transformations, this time to a Lagrangian where explicit \CP violation is possible.
From the analysis of the rather similar $\mathcal{L}_3$ Lagrangian, we can conclude that imposing $\mathcal{CP}_1$ to $\mathcal{L}_{3s}$ leads to $y_{10}=0$, and that imposing $\mathcal{CP}_2$ leads to $y_{01}=0$. As may have been expected, both make $I_{3s}$ vanish. This is a clear demonstration of the meaning of adding different \CP transformations - both lead to \CP conservation but not necessarily the same consequences. \footnote{Consider Yukawa-like couplings involving $Q$, $d^c$ and singlets $h_{01}$, $h_{10}$ and $h_{20}$ in the basis where these singlets correspond to $1_{01}$, $1_{10}$ and $1_{20}$ of $\Delta(27)$ respectively, then impose $\mathcal{CP}_1$ in that basis. The two couplings $y_{10}$ and $y_{20}$ are forced to vanish. Conversely, $\mathcal{CP}_2$ forces only one coupling to vanish, $y_{01}$. If one changes the basis it may be that e.g. $\mathcal{CP}_1$ is no longer associated with the $\Id$ matrix in the new basis, but it will nevertheless force two couplings to vanish whereas $\mathcal{CP}_2$ forces only one coupling to vanish. See \cite{Varzielas:2015fxa} for more specific considerations regarding basis changes.}

The type of CPI exemplified in $I_{3s}$ is very useful in the study of Yukawa-like terms between triplet, anti-triplet and singlets. For \CP to be conserved it is necessary that it vanishes.
In cases with with 3 $\Delta(27)$ singlets, it is also sufficient that the respective CPI vanishes for this type of Lagrangian to conserve \CP. The combinations of 3 singlets that automatically conserve \CP can be found \cite{Varzielas:2015fxa} to be 12 out of the total 84 combinations of 3 singlets. Curiously, these combinations are the 12 combinations of 3 singlets appearing in each of the 12 terms of $\mathcal{L}_{IX}$ in Eq.~\ref{Ls9} of Section \ref{singlets}.

To be clearer, consider explicitly the $y_1 h_{00} h_{01} h_{02}$ term in Eq.~\ref{Ls9}. We construct a Lagrangian similar to $\mathcal{L}_{3s}$ but with singlets $h_{00}$, $h_{01}$, $h_{02}$
\begin{equation}
\mathcal{L}_{3s_1} = y_{00} (L \nu^c)_{00} h_{00} + y_{01} (L \nu^c)_{02} h_{01} + y_{02} (L \nu^c)_{01} h_{02} + H.c.
\label{L3s1}
\end{equation}
then use Yukawa-like matrices $Y_{ij}$ and build the CPI similar to $I_{3s}$
\begin{equation}
I_{3s_1} \equiv \mathrm{Im} \Tr (Y_{00} Y_{01}^\dagger Y_{02} Y_{00}^\dagger Y_{01} Y_{02}^\dagger) \,.
\label{I3s1}
\end{equation}
Then we repeat these steps for the $y_2 h_{00} h_{10} h_{20}$ term in Eq.~\ref{Ls9} and so on. When inserting the conditions imposed by $\Delta(27)$ on the Yukawa-like matrices $Y_{ij}$, these 12 CPIs vanish automatically for any values of the Yukawa-like couplings
\begin{equation}
I_{3s_1} = I_{3s_2} = (...) = I_{3s_{12}} = 0 \,.
\end{equation}
One can explicitly build \CP transformations to prove that indeed the 12 respective Lagrangians $\mathcal{L}_{3s_1}$, $\mathcal{L}_{3s_2}$, $(...)$, $\mathcal{L}_{3s_{12}}$ automatically conserve \CP (for any values of the 3 Yukawa-like couplings of each respective Lagrangian).

As illustrated by $I_{3s}$ itself, the CPIs for the other 3 singlet combinations do not vanish automatically and appear with the respective $|y_{ij}|^2 |y_{kl}|^2 |y_{mn}|^2$, multiplied by a factor of $3 \omega$ or $3 \omega^2$. They are further examples of explicit geometrical \CP violation.

\subsection{Four or more singlets}

Any choice of 4 or more singlets will necessary include combinations of 3 that would allow \CP violation. For example, by adding any other singlet to the set $h_{00}$, $h_{01}$, $h_{02}$ in  $\mathcal{L}_{3s_1}$, we have a singlet with $h_{ij}$ with $i \neq 0$ and any of the $I_{3s}$-type CPIs involving $Y_{ij}$ with $Y_{00}$, $Y_{01}$, $Y_{02}$
\begin{align}
\mathrm{Im} \Tr (Y_{00} Y_{01}^\dagger Y_{ij} Y_{00}^\dagger Y_{01} Y_{ij}^\dagger) \,, \\
\mathrm{Im} \Tr (Y_{01} Y_{02}^\dagger Y_{ij} Y_{01}^\dagger Y_{02} Y_{ij}^\dagger) \,, \\
\mathrm{Im} \Tr (Y_{02} Y_{00}^\dagger Y_{ij} Y_{02}^\dagger Y_{00} Y_{ij}^\dagger) \,,
\end{align}
each of which falls in one of the 72 cases that are in general non-zero (unless the respective couplings are vanishing).

\section{Three triplets}
\label{three}

\subsection{Type II 2HDM with \secheadmath{\Delta(27)} triplet fermions \label{2HDM}}

We continue investigating Lagrangians with Yukawa-like terms between triplets and singlets, in the presence of 3 triplets of $\Delta(27)$. In contrast to the situation studied in \cite{deMedeirosVarzielas:2011zw, Varzielas:2012nn, Bhattacharyya:2012pi, Varzielas:2013sla, Varzielas:2013eta} where there is a scalar $\Delta(27)$ triplet coupling to fermions, we consider a situation where we generalise the $\mathcal{L}_3$ Lagrangian with additional anti-triplet $u^c$. The triplet $Q \sim 3$ now contains the SM quark $SU(2)$ doublets, and two anti-triplets $u^c, d^c \sim \bar{3}$ contain the up and down quark $SU(2)$ singlets.
The scalars are Higgs doublets $h_u \sim 1_{10}$ and $h_d \sim 1_{01}$ (we deviate here slightly from the notation used for scalars in other sections). With a $Z_2$ symmetry it is possible to have the $u^c$ couple only to $h_u$ and $d^c$ couple only to $h_d$, leading to a type II 2HDM (see e.g. \cite{Branco:2011iw}) where the actual Yukawa terms are constrained by $\Delta(27)$.
The Lagrangian is 
\begin{equation}
\mathcal{L}_{2HDM} =y_u (Q u^c)_{20} h_{u} + y_d (Q d^c)_{02} h_d + H.c. \,.
\label{L2HDM}
\end{equation}
We express the Lagrangian in terms of Yukawa matrices $Y_u$, $Y_d$ and apply the IA to this Lagrangian
\footnote{For simplicity, even though $Q$ and $d^c$ are explicitly fermions, we continue using the abridged notation of Eq.~\ref{scalarCP} rather than the more rigorous one of Eq.~\ref{fermionCP}.}
\begin{align}
h_u &\rightarrow e^{\ii p_u} h_u^* \,, \notag \\
h_d &\rightarrow e^{\ii p_d} h_d^* \,, \notag \\
Q &\rightarrow U_Q^T Q^* \,, \notag \\
u^c &\rightarrow U_u u^{c *} \,, \notag \\
d^c &\rightarrow U_d d^{c *}\,.
\label{2HDMCP}
\end{align}
The conditions on $Y_u$, $Y_d$ from imposing general \CP invariance on $\mathcal{L}_{2HDM}$ are
\begin{align}
U_Q Y_u U_u e^{\ii p_u}= Y_u^* \,,\\
U_Q Y_d U_d e^{\ii p_d}= Y_d^* \,.
\label{YuYd2HDM}
\end{align}
We see that we can without loss of generality redefine $U_u$ to absorb $e^{\ii p_u}$ and $U_d$ to absorb $e^{\ii p_d}$. We choose then $e^{\ii p_u}=e^{\ii p_d}=1$ and redefine $U_u$, $U_d$ accordingly.
\footnote{Note that we could not do this for $h_{10}$ and $h_{01}$ in $\mathcal{L}_3$ as there was only one anti-triplet $d^c$.}
In effect what this means is that we have, for $Y_u$ and $Y_d$, the same type of \CP conservation requirements that they would have in the SM. So we extrapolate from \cite{Bernabeu:1986fc}, rely on the Hermitian combinations $H_{u,d} \equiv Y_{u,d} Y_{u,d}^\dagger$ to eliminate the dependence in $U_{u,d}$
\begin{align}
U_Q H_u U_Q^\dagger= H_u^* \,,\\
U_Q H_d U_Q^\dagger= H_d^* \,,
\label{HuHd2HDM}
\end{align}
and conclude that
$\Tr \left[ H_u, H_d \right]^3 = 0$   \cite{Bernabeu:1986fc} is a necessary and sufficient condition for \CP conservation for $\mathcal{L}_{2HDM}$.
We can now use the $Y_u$ and $Y_d$ that $\Delta(27)$ would impose
\begin{equation}
Y_u=y_u
\begin{pmatrix}
1 & 0 & 0\\
0 & \omega^2 & 0\\
0 & 0 & \omega
\end{pmatrix} \,,
\end{equation}
which happens to correspond to the usual basis where $H_u$ is diagonal and
\begin{equation}
Y_d=y_d
\begin{pmatrix}
0 & 1 & 0\\
0 & 0 & 1\\
1 & 0 & 0
\end{pmatrix} \,.
\end{equation}
In the limit of unbroken $\Delta(27)$ in fact $H_u = H_d = \Id$ so it is clear that \CP is automatically conserved for any $y_u$, $y_d$.

Indeed, by using the IA on the Lagrangian we conclude that the only way to build CPIs is to keep Yukawa structures that couple to $d^c$ together, and also to keep Yukawa structures that couple to $u^c$ together, as that is the only way to cancel the respective $U_u$ and $U_d$ dependence. In addition to combining $H_u$ and $H_d$, one can use just $G_u$ or just $G_d$, but not mix both.

\subsection{Type II NHDM with $\Delta(27)$ triplet fermions \label{NHDM}}

Continuing from the Type II 2HDM Yukawa Lagrangian, using the IA it is relatively straightforward to generalise the conclusions for an increasing number of scalars. It is useful to classify each scalar and their respective Yukawa matrix according to their sector, i.e. an $h_{d_{ij}}$ couples to $d^c$ or an $h_{u_{kl}}$ couples to $u^c$. It is in general no longer possible to absorb the respective phases into $U_{u,d}$. The respective \CP conservation requirements include
\begin{align}
U_u^\dagger G_{u_{ij}} U_u= G_{u_{ij}}^* \,,\\
U_d^\dagger G_{d_{ij}} U_d= G_{d_{ij}}^* \,, \\
U_Q H_{u_{ij}} U_Q^\dagger= H_{u_{ij}}^* \,,\\
U_Q H_{d_{ij}} U_Q^\dagger= H_{d_{ij}}^* \,,
\end{align}
which cancel the dependence on phases $p_{u_{ij}}$, $p_{d_{ij}}$, but as we have seen already CPIs with the Hermitian combinations are automatically verified in unbroken $\Delta(27)$.

However, it is now possible to build CPIs of the type of $I_{3s}$ by using either 3 different $Y_{u_{ij}}$ or 3 different $Y_{d_{ij}}$ - without mixing the two sectors. Using the conclusions derived for the Lagrangians with 2 triplets (1 triplet and 1 anti-triplet) in Section \ref{2triplets}, we can extrapolate to this 3 triplet case (1 triplet and 2 anti-triplets).
Doing so, we conclude that each sector remains automatically \CP conserving when coupled to up to any 2 $\Delta(27)$ singlets, and can remain automatically \CP conserving when coupled to 3 $\Delta(27)$ singlets (if the 3 singlets are one of the 12 special combinations, as discussed in Section \ref{2triplets}).
If a specific representation is repeated in the up and down sector this still represents 2 fields, as they are distinguished by the type II $Z_2$ symmetry that distinguishes the sectors.
On the other hand, the 3 singlet representations chosen for the up sector and for the down sector need not be the same, so one can couple up to 6 different singlet representations without enabling \CP violation.
Conversely, coupling the triplets to 7 or more distinct singlet representations will not allow automatic \CP conservation, and the minimal singlet content that enables \CP violation is 3 singlet representations all in the same sector.

Recall also that even without additional symmetries, for $\Delta(27)$ triplet Weyl fermions one cannot construct $Q Q^\dagger h_{ij}$, $d^c d^{c \dagger} h_{kl}$ or $u^c u^{c \dagger} h_{mn}$ due to Lorentz invariance (even though the combinations would be $\Delta(27)$ invariant).

If one continues to generalise this class of Lagrangians to 4 or more triplets the conclusion is indeed that we can separately treat each Yukawa-like sector (a distinct triplet to anti-triplet pairing). This is because the general \CP matrices appearing for each sector are unrelated, and that constrains the types of CPIs that can be built in the IA.
In $\Delta(27)$, each sector can couple to as many as 3 different singlets before \CP violation arises as a possibility. This extends the conclusion derived in Section \ref{2triplets} for Lagrangians with a single Yukawa-like sector, $\mathcal{L}_{3}$, $\mathcal{L}_{3s}$ and the 12 special combinations $\mathcal{L}_{3s_1}$, $(...)$.

It is interesting to note that with 3 sectors, which arises as a possibility with 4 or more triplets, the full representation content of $\Delta(27)$ with all 9 singlets can be present while \CP is still automatically conserved. Given that cases with 4 or more triplets no longer have a counterpart with the SM quarks, we relegate a more detailed analysis to appendix \ref{sec:4more}. In addition we present some examples of specific \CP transformations in appendix \ref{spe}, which includes an existence proof of \CP transformations for a Lagrangian with all the irreducible representations of $\Delta(27)$.

\section{Conclusions \label{Conclusions}}

The group $\Delta(27)$ is very interesting from the point of view of \CP properties. In this work we considered several $\Delta(27)$ invariant Lagrangians with Yukawa-like terms (tri-linears) and studied them with the invariant approach.

Our dual purpose was to demonstrate the usefulness of the invariant approach in Lagrangians invariant under discrete family symmetries, and simultaneously to explore the \CP properties of $\Delta(27)$. The method is independent of the group when the \CP-odd invariants are constructed, and the group details are needed only to obtain coupling matrices in some convenient basis, which can then be used in the \CP-odd invariants to obtain basis-independent quantities signalling \CP violation.

Starting with simple cases where the field content includes only 1-dimensional representations of $\Delta(27)$ (singlets), the invariant approach reveals what are the relevant physical phases, which turn out to be specific relative phases of the complex couplings.

We then turned to consider Yukawa-like terms involving $\Delta(27)$ triplet and anti-triplet, starting with a single 3-dimensional representation (triplet) and progressing to two and more triplets, where it becomes helpful to refer to sectors of distinct pairs of triplet and anti-triplet.

The conclusions derived for the two triplet case with one sector are that \CP is automatically conserved for Yukawa-like terms involving up to any 2 $\Delta(27)$ singlets and for 12 special combinations out the total 84 combinations of 3 singlets. The other cases are examples of explicit geometrical \CP violation.

Based on these results, the invariant approach allows us to extrapolate for cases with three or more triplets. The same type of conclusion holds independently for each sector, and therefore with 3 sectors it is even possible to have all 9 $\Delta(27)$ singlets present while automatically conserving \CP.

We have therefore completed a fairly exhaustive analysis of unbroken $\Delta(27)$ Lagrangians. 
The analysis here should provide a useful guide
for formulating future realistic models in which $\Delta(27)$ is spontaneously broken.
However the main motivation was to highlight the utility and power of the invariant approach for Lagrangians with discrete family symmetries.

\section*{Acknowledgments}
This project has received funding from the European Union's Seventh Framework Programme for research, technological development and demonstration under grant agreement no PIEF-GA-2012-327195 SIFT.
The authors also acknowledge partial support from the European Union FP7 ITN-INVISIBLES (Marie Curie Actions, PITN- GA-2011- 289442),
and by Funda\c{c}\~{a}o para a Ci\^{e}ncia e a Tecnologia
(FCT, Portugal) through the projects CERN/FP/123580/2011, PTDC/FIS-NUC/0548/2012 and CFTP-FCT Unit 777 (UID/FIS/00777/2013) which are partially funded through
POCTI (FEDER), COMPETE, QREN and EU.
We thank CERN for hospitality.

\appendix

\section{Four or more triplets \label{sec:4more}}

The final generalisation that we consider is to add more fields transforming as triplet representations.
Following from the three triplet case with triplet $Q$ and anti-triplets $d^c$, $u^c$, we add another anti-triplet $x^c$. We continue to assume that each anti-triplet has its own sector of $\Delta(27)$ singlets denoted as $h_{d_{ij}}$, $h_{u_{kl}}$, $h_{x_{mn}}$ due to e.g. an Abelian symmetry. Using the IA and considering how CPIs can be constructed we extend the previous results to conclude that the relevant CPIs are of $I_{3s}$ type for each sector, due to the different $U_d$, $U_u$, $U_x$ matrices
\begin{equation}
\mathcal{L}_{4Q} =Y_{d_{ij}} (Q d^c) h_{d_{ij}} + Y_{u_{kl}} (Q u^c) h_{u_{kl}} + Y_{x_{mn}} (Q x^c) h_{x_{mn}}  +  H.c. \,,
\label{L_4Q}
\end{equation}
\begin{align}
Q &\rightarrow U_Q^T Q^* \,, \notag \\
d^c &\rightarrow U_d d^{c *} \,, \notag \\
u^c &\rightarrow U_u u^{c *} \,, \notag \\
x^c &\rightarrow U_x x^{c *} \,.
\label{4Q}
\end{align}

It is interesting that at 4 triplets (in this case 1 triplet and 3 anti-triplets) we have reached a situation where \CP can be automatically conserved even with fields transforming as each of the 9 $\Delta(27)$ singlets (one example is $h_{d_{00}}$, $h_{d_{01}}$, $h_{d_{02}}$, $h_{u_{10}}$, $h_{u_{11}}$, $h_{u_{12}}$, $h_{x_{20}}$, $h_{x_{21}}$, $h_{x_{22}}$).

At this stage it should be relatively clear that adding further anti-triplets $y^c$, $z^c$, $(...)$, with their own singlet sector, the IA still shows the relevant CPIs to be of $I_{3s}$ type for each sector because of the different $U_y$, $U_z$, $(...)$.

Conversely, the IA also allows to generalise to cases where there are multiple triplets and multiple anti-triplets. Considering distinct triplets $Q$, $L$, and several anti-triplets, in general the conclusion depends on how many separate sectors are present, but there are some subtleties.
For example, take the Lagrangian 
\begin{equation}
\mathcal{L}_{4} =Y_{d_{ij}} (Q d^c) h_{d_{ij}} + Y_{u_{kl}} (Q u^c) h_{u_{kl}} + Y_{e_{mn}} (L e^c) h_{e_{mn}} \,,
\label{L_4L}
\end{equation}
with the usual general transformations for the singlets each with their own phase and
\begin{align}
L &\rightarrow U_L^T L^* \,, \notag \\
Q &\rightarrow U_Q^T Q^* \,, \notag \\
d^c &\rightarrow U_d d^{c *} \,, \notag \\
u^c &\rightarrow U_u u^{c *} \,, \notag \\
e^c &\rightarrow U_e e^{c *} \,.
\label{4L}
\end{align}
In this case there are 3 sectors like in $\mathcal{L}_{4Q}$, because some additional symmetry distinguishes triplets $Q$ and $L$ such that $L$ pairs only with $e^c$, $h_{e_{mn}}$.
Consider instead a situation where $Q$ and $L$ couple to the same anti-triplets and singlets, like in the Lagrangian
\begin{align}
\mathcal{L}_{4QL}= & Y_{Qd_{ij}} (Q d^c) h_{d_{ij}} + Y_{Qu_{kl}} (Q u^c) h_{u_{kl}} + Y_{Qe_{mn}} (Q e^c) h_{e_{mn}} \label{L_4QL} \\ 
+ & Y_{Ld_{ij}} (L d^c) h_{d_{ij}} + Y_{Lu_{kl}} (L u^c) h_{u_{kl}} + Y_{Le_{mn}} (L e^c) h_{e_{mn}} +  H.c. \,, \notag
\end{align}
which has 6 sectors. Note though that while e.g. $Q d^c$ and $L d^c$ appear to be distinct sectors, if $Q d^c$ has automatic \CP conservation having up to 3 $h_{d_{ij}}$ singlets, this applies also to $L d^c$, because the singlets $h_{d_{ij}}$ are the same. 

\section{Specific \CP matrices \label{spe}}

It is interesting to compare the IA to the construction of specific \CP transformations for Lagrangians of $\mathcal{L}_3$ type with any of the singlets
\begin{equation}
\mathcal{L} = Q Y_{ij} d^c h_{ij} +  (...) + H.c. \,,
\label{L3type}
\end{equation}
\begin{align}
h_{ij} &\rightarrow e^{\ii p_{ij}} h_{ij}^* \,, \notag \\
Q &\rightarrow U_Q^T Q^* \notag \,, \\
d^c &\rightarrow U_d d^{c *} \,.
\label{L3typeCP}
\end{align}

The matrices associated to each singlet $h_{ij}$ are, due to $\Delta(27)$
\begin{equation}
Y_{00}=y_{00}
\begin{pmatrix}
1 & 0 & 0\\
0 & 1 & 0\\
0 & 0 & 1
\end{pmatrix} \,,
\end{equation}
\begin{equation}
Y_{01}=y_{01}
\begin{pmatrix}
0 & 1 & 0\\
0 & 0 & 1\\
1 & 0 & 0
\end{pmatrix} \,,
\end{equation}
\begin{equation}
Y_{02}=y_{02}
\begin{pmatrix}
0 & 0 & 1\\
1 & 0 & 0\\
0 & 1 & 0
\end{pmatrix} \,,
\end{equation}
\begin{equation}
Y_{10}=y_{10}
\begin{pmatrix}
1 & 0 & 0\\
0 & \omega & 0\\
0 & 0 & \omega^2
\end{pmatrix} \,,
\end{equation}
\begin{equation}
Y_{11}=y_{11}
\begin{pmatrix}
0 & \omega & 0\\
0 & 0 & \omega^2\\
1 & 0 & 0
\end{pmatrix} \,,
\end{equation}
\begin{equation}
Y_{12}=y_{12}
\begin{pmatrix}
0 & 0 & \omega^2 \\
1 & 0 & 0\\
0 & \omega & 0
\end{pmatrix} \,,
\end{equation}
\begin{equation}
Y_{20}=y_{20}
\begin{pmatrix}
1 & 0 & 0\\
0 & \omega^2 & 0\\
0 & 0 & \omega
\end{pmatrix} \,,
\end{equation}
\begin{equation}
Y_{21}=y_{01}
\begin{pmatrix}
0 & \omega^2 & 0\\
0 & 0 & \omega\\
1 & 0 & 0
\end{pmatrix} \,,
\end{equation}
\begin{equation}
Y_{22}=y_{02}
\begin{pmatrix}
0 & 0 & \omega\\
1 & 0 & 0\\
0 & \omega^2 & 0
\end{pmatrix} \,.
\end{equation}

For each Yukawa-like term $Q Y_{ij} d^c h_{ij}$ we have a \CP conservation requirement on the Yukawa-like matrix $Y_{ij}$
\begin{equation}
U_Q Y_{ij} U_d e^{\ii p_{ij}}= Y_{ij}^* \,,
\end{equation}
therefore in general the $U_Q$ and $U_d$ matrices that respect this requirement are different for each $Y_{ij}$. Within this point of view, the possibility of \CP violation arises when it is impossible to have even a single set of $U_Q$, $U_d$ and $p_{ij}$ transformations that simultaneously fulfil the distinct requirements of all $Y_{ij}$ that are present in the Lagrangian.
Checking this can be quite laborious as it should be done with complete generality and indeed one of the advantages of the IA is that usually one need not check for the existence of such transformations.
For illustration purposes we fix for simplicity the respective $p_{ij} = e^{- 2 \ii \arg[y_{ij}] }$, $U_Q = \Id$, and $U_d$ to be diagonal. For these choices we present for each $Y_{ij}$ the respective $U_d$
\begin{equation}
U_d=
\begin{pmatrix}
1 & 0 & 0\\
0 & 1 & 0\\
0 & 0 & 1
\end{pmatrix} \,,
\end{equation}
corresponds to the requirement of the $h_{00}$, $h_{01}$ and $h_{02}$ Yukawa-like coupling;
\begin{equation}
U_d=
\begin{pmatrix}
1 & 0 & 0\\
0 & \omega & 0\\
0 & 0 & \omega^2
\end{pmatrix} \,,
\label{Ud10}
\end{equation}
corresponds to the requirement of the $h_{10}$, $h_{11}$ and $h_{12}$ Yukawa-like coupling;
\begin{equation}
U_d=
\begin{pmatrix}
1 & 0 & 0\\
0 & \omega^2 & 0\\
0 & 0 & \omega
\end{pmatrix} \,,
\label{Ud20}
\end{equation}
corresponds to the requirement of the $h_{20}$, $h_{21}$ and $h_{22}$ Yukawa-like coupling.
These \CP transformations are with loss of generality (e.g. $U_d$ need not be diagonal), but they are still an existence proof of a valid \CP transformation for each singlet by itself, and for the specific groups of 3 singlets shown. It naturally agrees with what was obtained through the IA for the more general case.
For example, we knew already that with a single sector $Q d^c$, these 3 combinations of three singlets belong to the 12 that automatically conserve \CP.

Furthermore, with 3 sectors $Q d^c$, $Q u^c$ and $Q x^c$ we have in addition to $U_d$ also $U_u$ and $U_x$, enabling the possibility to have all 9 $\Delta(27)$ singlets in Yukawa-like terms with triplets while automatically conserving \CP for any arbitrary (non-zero) complex value of the nine $y_{ij}$. In this 3 sector case with $\mathcal{L}_{4Q}$ in Eq.~\ref{L_4Q}, an existence proof for $h_{d_{00}}$, $h_{d_{01}}$, $h_{d_{02}}$, $h_{u_{10}}$, $h_{u_{11}}$, $h_{u_{12}}$, $h_{x_{20}}$, $h_{x_{21}}$, $h_{x_{22}}$ follows by keeping $p_{ij} = e^{- 2 \ii \arg[y_{ij}] }$, $U_Q = \Id$, $U_d = \Id$ and taking $U_u$ and $U_x$ respectively as the diagonal matrices appearing in Eq.~\ref{Ud10},\ref{Ud20}.


\begin{thebibliography}{27}

\bibitem{Bernabeu:1986fc}
  J.~Bernabeu, G.~C.~Branco and M.~Gronau,
  Phys.\ Lett.\ B {\bf 169} (1986) 243.

\bibitem{Grimus:1995zi}
  W.~Grimus and M.~N.~Rebelo,
  Phys.\ Rept.\  {\bf 281} (1997) 239
  [hep-ph/9506272].

\bibitem{Ecker:1981wv}
  G.~Ecker, W.~Grimus and W.~Konetschny,
  Nucl.\ Phys.\ B {\bf 191} (1981) 465;
  G.~Ecker, W.~Grimus and H.~Neufeld,
  Nucl.\ Phys.\ B {\bf 247} (1984) 70.

\bibitem{Jarlskog:1985ht}
  C.~Jarlskog,
  Phys.\ Rev.\ Lett.\  {\bf 55} (1985) 1039.

\bibitem{Branco:1986gr} 
  G.~C.~Branco, L.~Lavoura and M.~N.~Rebelo,
  Phys.\ Lett.\ B {\bf 180}, 264 (1986);
  G.~C.~Branco, M.~N.~Rebelo and J.~I.~Silva-Marcos,
  Phys.\ Rev.\ Lett.\  {\bf 82}, 683 (1999)
  [hep-ph/9810328];
  H.~K.~Dreiner, J.~S.~Kim, O.~Lebedev and M.~Thormeier,
  Phys.\ Rev.\ D {\bf 76}, 015006 (2007)
  [hep-ph/0703074 [HEP-PH]].

\bibitem{Ade:2015xua}
  P.~A.~R.~Ade {\it et al.}  [Planck Collaboration],
  arXiv:1502.01589 [astro-ph.CO],
  R.~H.~Cyburt, B.~D.~Fields, K.~A.~Olive and T.~H.~Yeh,
  arXiv:1505.01076 [astro-ph.CO].

\bibitem{Branco:2015hea}
  G.~C.~Branco, I.~de Medeiros Varzielas and S.~F.~King,
  arXiv:1502.03105 [hep-ph].

\bibitem{Varzielas:2015fxa}
  I.~de~Medeiros~Varzielas,
  arXiv:1503.02633 [hep-ph].

\bibitem{Branco:2011iw}
  G.~C.~Branco, P.~M.~Ferreira, L.~Lavoura, M.~N.~Rebelo, M.~Sher and J.~P.~Silva,
  Phys.\ Rept.\  {\bf 516} (2012) 1
  [arXiv:1106.0034 [hep-ph]].

\bibitem{Feruglio:2012cw}
  F.~Feruglio, C.~Hagedorn and R.~Ziegler,
  JHEP {\bf 1307} (2013) 027
  [arXiv:1211.5560 [hep-ph]].
  
\bibitem{Holthausen:2012dk}
  M.~Holthausen, M.~Lindner and M.~A.~Schmidt,
  JHEP {\bf 1304} (2013) 122
  [arXiv:1211.6953 [hep-ph]].

\bibitem{Branco:1983tn}
  G.~C.~Branco, J.~M.~Gerard and W.~Grimus,
  Phys.\ Lett.\ B {\bf 136} (1984) 383.

\bibitem{deMedeirosVarzielas:2006fc}
  I.~de Medeiros Varzielas, S.~F.~King and G.~G.~Ross,
  Phys.\ Lett.\ B {\bf 648} (2007) 201
  [hep-ph/0607045].

\bibitem{Ma:2006ip}
  E.~Ma,
  Mod.\ Phys.\ Lett.\ A {\bf 21} (2006) 1917
  [hep-ph/0607056].

\bibitem{Luhn:2007uq}
  C.~Luhn, S.~Nasri and P.~Ramond,
  J.\ Math.\ Phys.\  {\bf 48} (2007) 073501
  [hep-th/0701188].

\bibitem{Ishimori:2010au}
  H.~Ishimori, T.~Kobayashi, H.~Ohki, Y.~Shimizu, H.~Okada and M.~Tanimoto,
  Prog.\ Theor.\ Phys.\ Suppl.\  {\bf 183} (2010) 1
  [arXiv:1003.3552 [hep-th]].

\bibitem{deMedeirosVarzielas:2011zw}
  I.~de Medeiros Varzielas and D.~Emmanuel-Costa,
  Phys.\ Rev.\ D {\bf 84} (2011) 117901
  [arXiv:1106.5477 [hep-ph]].

\bibitem{Varzielas:2012nn}
  I.~de Medeiros Varzielas, D.~Emmanuel-Costa and P.~Leser,
  Phys.\ Lett.\ B {\bf 716} (2012) 193
  [arXiv:1204.3633 [hep-ph]].

\bibitem{Bhattacharyya:2012pi}
  G.~Bhattacharyya, I.~de Medeiros Varzielas and P.~Leser,
  Phys.\ Rev.\ Lett.\  {\bf 109} (2012) 241603
  [arXiv:1210.0545 [hep-ph]].

\bibitem{Varzielas:2013sla}
  I.~de Medeiros Varzielas and D.~Pidt,
  J.\ Phys.\ G {\bf 41} (2014) 025004
  [arXiv:1307.0711 [hep-ph]].

\bibitem{Varzielas:2013eta}
  I.~Medeiros Varzielas and D.~Pidt,
  JHEP {\bf 1311} (2013) 206
  [arXiv:1307.6545 [hep-ph], arXiv:1307.6545].

\bibitem{Chen:2014tpa}
  M.~C.~Chen, M.~Fallbacher, K.~T.~Mahanthappa, M.~Ratz and A.~Trautner,
  Nucl.\ Phys.\ B {\bf 883} (2014) 267
  [arXiv:1402.0507 [hep-ph]].

\bibitem{Ivanov:2014doa}
  I.~P.~Ivanov and C.~C.~Nishi,
  JHEP {\bf 1501} (2015) 021
  [arXiv:1410.6139 [hep-ph]].

\bibitem{Fallbacher:2015rea}
  M.~Fallbacher and A.~Trautner,
  Nucl.\ Phys.\ B {\bf 894} (2015) 136
  [arXiv:1502.01829 [hep-ph]].

\bibitem{Varzielas:2012pd}
  I.~de Medeiros Varzielas,
  JHEP {\bf 1208} (2012) 055
  [arXiv:1205.3780 [hep-ph]].

\bibitem{Ivanov:2013nla}
  I.~P.~Ivanov and L.~Lavoura,
  Eur.\ Phys.\ J.\ C {\bf 73} (2013) 4,  2416
  [arXiv:1302.3656 [hep-ph]].

\bibitem{Varzielas:2013zbp}
  I.~de Medeiros Varzielas,
  J.\ Phys.\ Conf.\ Ser.\  {\bf 447} (2013) 012038
  [arXiv:1302.3991 [hep-ph]].

\bibitem{Nishi:2013jqa}
  C.~C.~Nishi,
  Phys.\ Rev.\ D {\bf 88} (2013) 3,  033010
  [arXiv:1306.0877 [hep-ph]].

\end{thebibliography}
\end{document}